# Angular Distribution of Electrons in Photoionization of Atoms Adsorbed on a Graphene Sheet

A. S. Baltenkov and V. A. Pikhut

Institute of Ion-Plasma and Laser Technologies, Uzbek Academy of Sciences
100125, Tashkent, Uzbekistan

arkbalt@mail.ru; verapi@mail.ru

*Abstract*

Within the framework of a model representing the potential of a graphene sheet $U(z)$ as an electro-neutral layer formed by smeared carbon atoms, the effect of this potential on spectral characteristics of atoms adsorbed on a graphene sheet has been studied. Since the distance between the adsorbed atom nucleus and sheet surface significantly exceeds the radii of inner atomic shells the potential $U(z)$ makes influence on the continuum wave functions only. Their behavior in the upper semi-space ($z>0$) and in the lower one ($z<0$) where the adsorbed atom is located is defined by a jump of the logarithmic derivative of the wave function for $z=0$. The photoelectron angular distributions have been calculated for different mutual positions of the polarization vector **e** and the axis Z normal to the sheet surface. It has been shown that the existence of the electron waves reflected from the potential $U(z)$ leads to evident asymmetry of the angular distribution relative to the plane $z=0$. The experimental observation of this effect is of great interest for photoelectron spectroscopy of atoms localized on graphene structures.

## Introduction

The number of papers devoted to investigation of carbon nanostructures and their feasible applications has seen rapid increases (see Lehtinen, P. O., et al 2003; Saito, R., et al 1992; Ajayan, P.M. 1999; Gerber, I. C., et al 2010; Xia, H., et al 2008; Mishra, K. N., et al 2012; Rangel, N. L. and J. M. Seminario. 2010) and references therein). The atomic-scale understanding of the properties of the adatoms on a graphene sheet has become an essential part of such studies. The photoelectron spectroscopy is an effective precise method to investigate the electron structure of adatoms (Carlson, T.A. 1975). The study of the interaction processes of ionizing radiation with single adatoms and surface structures on their basis will not only provide an answer to a number of the intriguing fundamental scientific problems of atomic, molecular physics and surface physics but also promote the development of surface-structures nano-engineering.

In this paper we study the angular distribution of photoelectrons knocked-out from the deep subshells of atoms adsorbed on a graphene sheet. Since we deal with physical adsorption when the adatom electronic structure is not subject to significant changes the electronic subsystems of graphene sheet and adatom can be considered independently. The distance $d$ between the adsorbed atom nucleus and sheet surface is several atomic units (1.87Å (Lehtinen, P. O., et al 2003)) and so it significantly exceeds the radii of inner atomic shells. That allows considering the wave functions of these shells with great accuracy as coinciding with the wave functions of the free atom. As far as the continuum wave functions are concerned, for near-threshold photoionization the wavelength of electrons produced in this process exceeds by far the bond lengths between the carbon atoms in the graphene sheet. This effectively obliterates the influence of the individual carbon atoms on the knocked-out low-energy photoelectron, thereby making the phenomenological potential $U(z)$ formed by the $z$=0-plane-smeared carbon atoms to be an excellent approximation. A matrix element of the optical transition is formed near the adatom nucleus where the potential of graphene sheet is equal to zero. Therefore the specific shape of the function $U(z)$ has no influence on the photoionization amplitude. This allows the real interaction potential $U(z)$ having a finite thickness to be replaced by a zero-range potential $U(z) = -A\delta(z)$ providing a necessary jump of the logarithmic derivative of the wave function at $z=0$.

A similar method based on the so-called Dirac-bubble-potential $U(r) = -A\delta(r - R)$ (Lohr, L. L. et al 1992) for the $C_{60}$ fullerene shell was developed in (Baltenkov, A. S. 1999; Dolmatov, V. K., et al 2004; Baltenkov, A. S., et al 2010) where this potential was used to describe the photoionization of endohedral atoms M@$C_{60}$ as well. As a matter of fact, the M atom photoionization on a graphene sheet is the photoionization of the same atom M in the system M@$C_{60}$ where the fullerene shell is expanded on a plane. The difference between the structures of graphene formed by 6-rings carbon atoms and the fullerene shell consisting of 5- and 6-rings plays no principal role.

Introducing the $\delta$ - potential in the wave equation, *per se*, is a suitable mathematical way allowing replacement of the real interaction potential having a finite radius of action by a zero-range potential providing the necessary boundary conditions for the wave function of a particle moving in this potential. This idea originates from the classical paper on deuteron photodetachment (Bethe, H., et al 1935). This way allows avoiding arbitrariness for selecting the parameters of the potential well $U(z)$. This arbitrariness is restricted by a single parameter $A$ defined from experiment. It is evident that, as any model description, the $\delta$ - potential model has a limited field of application. Within this model it is impossible to consider the processes for which the electronic structure of the carbon nanostructures plays an essential role. Such processes are, in particular, those of $C_{60}$ molecule photoionization. For them critically important are the geometrical properties of the fullerene shell, such as its diameter and thickness (Madjet, M. E., et al 2008; Rüdel, A., et al 2002). The fullerene cage cannot be also considered as a source of the static potential for description of photoionization of atom M endohedrally confined in molecule $C_{60}$. For correct description of this process for some photon energies it is necessary to use the more sophisticated approaches for the fullerene electronic hull (McCune, M. A., et al 2009).

The sequences of introducing into consideration of the atom photoionization processes the carbon nano-structures potentials $U(r) = -A\delta(r - R)$ and $U(z) = -A\delta(z)$ are principally different, which is understandable already at the qualitative level. In the first case the spherically symmetric potential leads to radical changes in the photoionization *total cross sections* where the so-called *confinement resonances* discovered in the recent experiments (Kilcoyne, A.L.D., et al 2010) appear. The amplitude of these resonances relate to the number of backscattered waves emitted by the nearby C atoms equidistantly located from the M atom (Baltenkov, A. S., et al 2010) . At the same time the photoelectron angular distributions of endohedral atoms M@$C_{60}$ remain unchanged, i.e. they are described by the same linear combinations of trigonometric functions as in the case of the free atom M; there are only changes in the coefficients at these function. In the second case because of non-sphericity of the potential $U(z)$ the principal changes should be expected, first of all, in the *photoelectron angular distributions* while the total cross sections will be slightly distorted owing to the small (as compared to $C_{60}$) coordination number, i.e. the number of carbon atoms equidistantly located from the adatom nucleus. However the question whether the changes in the shape of the angular distributions are observable and what their character is can be resolved by calculating the differential cross sections. In this paper we will obtain an answer to these questions by considering the photoionization of atom near the graphene sheet within the framework of a simple model.

## DIFFERENTIAL PHOTOIONIZATION CROSS SECTION

In the spherical coordinate system connected with the sheet surface the axis Z is considered to be directed along the normal to the surface. The ionized atom M is placed in the lower semi-space (Fig. 1).

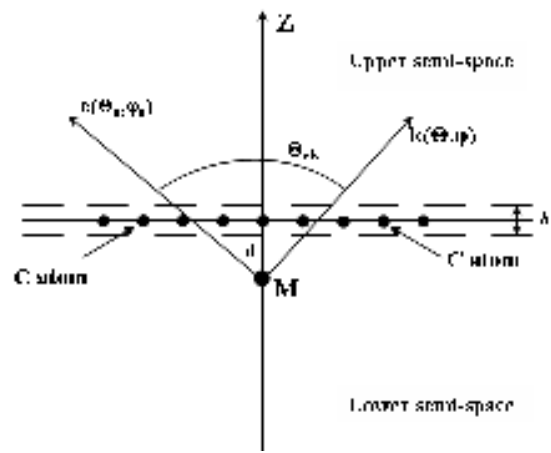

FIG. 1 THE SCHEME OF THE ADSORBED M ATOM LOCATION ON THE GRAPHENE SHEET. THE PLANE WHERE THE ATOM CARBONS ARE LOCATED IS PERPENDICULAR TO THE PAGE

The spherical coordinates of the radiation polarization vector **e** are $(\vartheta_e, \varphi_e)$ and the coordinates of the vector of photoelectron momentum **k** are $(k, \vartheta, \varphi)$. It is known that the differential cross section of dipole photoionization of atom for linearly polarized light is determined by the cosine of the angle $\vartheta_{ek}$ (between the vectors **e** and **k**) and has the general form (Bethe, H. 1933)

$$\frac{d\sigma}{d\Omega} = \frac{\sigma}{4\pi}[1 + \beta P_2(\cos\vartheta_{ek})]. \qquad (1)$$

Here $\sigma$ is the total photoionization cross section; $\beta$ is the dipole asymmetry parameter. In the above-described coordinate system the argument of the Legendre polynomial $P_2(\cos\vartheta_{ek})$ is represented as

$$\cos\vartheta_{ek} = \cos\vartheta_e \cos\vartheta + \sin\vartheta_e \sin\vartheta \cos(\varphi_e - \varphi). \qquad (2)$$

For the fixed coordinates of the radiation polarization vector $\vartheta_e$ and $\varphi_e$ the shape of the electron angular distribution in the coordinate system connected with the sheet surface we will define as the following function of spherical angles $\vartheta$ and $\varphi$

$$F(\vartheta, \varphi) = \frac{[1 + \beta P_2(\cos\vartheta_{ek})]}{1 + \beta}. \qquad (3)$$

Since we are interested in the shape of the angular distribution only, we normalize this function so that it is equal to unity for $\vartheta_{ek} = 0$. The vector $\mathbf{J} = \mathbf{k}F(\vartheta,\varphi)$ is a flux of photoelectrons knocked-out by the ionizing atom (here and throughout the text the atomic system of units $e = m = \hbar = 1$ is used). The Cartesian components of this flux are defined by the following formulas

$$j_x = k_x F(\vartheta,\varphi) = k F(\vartheta,\varphi)\sin\vartheta\cos\varphi,$$
$$j_y = k_y F(\vartheta,\varphi) = k F(\vartheta,\varphi)\sin\vartheta\sin\varphi, \qquad (4)$$
$$j_z = k_z F(\vartheta,\varphi) = k F(\vartheta,\varphi)\cos\vartheta.$$

The photoelectron wave function that corresponds to the flux **J** we represent as

$$\psi_{\mathbf{k}}(\mathbf{r}) = F(\vartheta,\varphi)e^{i\mathbf{k}\cdot\mathbf{r}}. \qquad (5)$$

Writing down the electron wave function in the form (5), we consider electron as a free particle neglecting by the Coulomb interaction of photoelectron with the atomic residue since the Coulomb field gives for the electron flux density the corrections that go to zero at $r \to \infty$ (Landau, L. D. and E. M. Lifshitz 1965).

## GRAPHENE SHEET EFFECT

Until now the above-given formulas relate to the free M atom. In the case of the ionization of adsorbed atom, the potential plane formed by the graphene sheet distorts the final-state wave function (5). Near the photoeffect threshold when the photoelectron wave length $\lambda$ significantly exceeds the distance between the carbon atoms in the 6-rings forming the graphene sheet its potential (by analogy with the Dirac-bubble-potential for $C_{60}$ shell (Lohr, L. L. et al 1992)) we represent as a phenomenological potential $U(z) = -A\delta(z)$ simulating the smeared carbon atoms.

Within this model description the wave function of photoelectron leaving the atomic residue will obey the following Schrödinger equation

$$\Delta\psi_{\mathbf{k}} + 2[E - U(z)]\psi_{\mathbf{k}} = 0;$$
$$E = k^2/2 = (k_x^2 + k_y^2 + k_z^2)/2. \qquad (6)$$

The solution of this equation is written as

$$\psi_{\mathbf{k}}(\mathbf{r}) = F(\vartheta,\varphi)e^{ik_x x + ik_y y}\varphi(z). \qquad (7)$$

It is evident that for $U(z) \equiv 0$ the function (7) coincides, as it should be, with the function (5). For the function $\varphi(z)$ we have the equation

$$\frac{\partial^2\varphi}{\partial z^2} + [k_z^2 + 2A\delta(z)]\varphi(z) = 0. \qquad (8)$$

The solution of Eq.(8) we write in the form (Flugge, S. 1971)

$$\varphi(z) = e^{ik_z z} + Be^{-ik_z z} \quad \text{for } z<0,$$
$$\varphi(z) = (1+D)e^{ik_z z} \quad \text{for } z>0. \qquad (9)$$

In the formulas (9) we distinguish three waves: an incident wave with unity intensity, a reflected wave with intensity $|B|^2$ and a wave passed over the potential well with intensity $|1+D|^2$. Because of continuity of the function $\varphi(z)$ at $z=0$ we obtain for the scattering amplitudes the following relation: $B=D$. Integrating the equation (8) over $z$ within the limits $\pm\varepsilon \to 0$, we obtain the equation connecting a jump of the logarithmic derivative of the wave function $\Delta L(z = 0)$ with the strength of the delta-potential:

$$\Delta L = \left[\frac{\partial\varphi}{\partial z}\Big|_{+\varepsilon} - \frac{\partial\varphi}{\partial z}\Big|_{-\varepsilon}\right]\frac{1}{\varphi(0)} = -2A. \qquad (10)$$

Substituting the functions (9) in Eq.(10), we obtain the following expression for the scattering amplitudes

$$B = D = -\frac{A}{A + ik_z}. \quad (11)$$

Owing to electron wave reflection from the potential plane the electron flux vector has now the following z-component in the lower semi-space

$$j_z^l = F(\vartheta,\varphi)[1+|B|^2]k\cos\vartheta, \quad (12)$$

and in the upper one

$$j_z^u = F(\vartheta,\varphi)|1+D|^2 k\cos\vartheta = [1-|B|^2]k\cos\vartheta. \quad (13)$$

As far as the Cartesian components of the photoelectron flux $j_x$ and $j_y$ are concerned, they are undistorted by the potential $U(z)$ and defined as before by the formulas (4). Finally, the formulas associating the function $F(\vartheta,\varphi)$ for the free atom with the angular distribution of electrons knocked out of the atom localized near the potential well $U(z)$ in the Cartesian coordinates have the following form

$$S_x = F(\vartheta,\varphi)\sin\vartheta\cos\varphi,$$

$$S_y = F(\vartheta,\varphi)\sin\vartheta\sin\varphi, \quad (14)$$

$$S_z = F(\vartheta,\varphi)[1-|B|^2]\cos\vartheta, \text{ for the upper semi-space,}$$

$$S_z = F(\vartheta,\varphi)[1+|B|^2]\cos\vartheta, \text{ for the lower semi-space.}$$

The function $|B|^2$ in (14) has the form

$$|B|^2 = \frac{A^2}{A^2 + k^2\cos^2\vartheta}. \quad (15)$$

From (15) it follows that when the electron moves from the upper semi-space to the lower one or vice versa, i.e. at angles $\vartheta = \pi/2$ or $3\pi/2$, the z-components of the function $S_z$ Eq.(14) have the jump due to partial reflection of the electron flux from the potential well.

Let us investigate how the derived formulas are modified if the finite thickness of the graphene film is taken into consideration. The formulas (9) will have now the following form

$$\varphi(z) = e^{ik_z z} + Be^{-ik_z z} \quad \text{for } z < -b/2,$$

$$\varphi(z) = A\xi(z) + C\zeta(z) \quad \text{for } b/2 > z > -b/2, \quad (16)$$

$$\varphi(z) = (1+D)e^{ik_z z} \quad \text{for } z > b/2.$$

Here $A\xi(z) + C\zeta(z)$ is the linear combination of the regular and irregular solutions of the Schrödinger equation inside the graphene sheet. Matching the logarithmic derivatives at the points $z = \pm b/2$ (here $b$ is the effective film thickness shown in Fig. 1 by the dashed lines), we obtain for the amplitude of the reflected wave the following expression

$$|B|^2 = \left|\frac{L_- - L_+}{L_- + L_+}\right|^2. \quad (17)$$

Here $L_+$ and $L_-$ are the logarithmic derivatives at the points $z = \pm b/2$, i.e. the two model parameters to be defined. Let us suppose that the potential $U(z)$ is a squared well with the depth $W$ and the thickness $b$. Then the linear combination of the solutions within the range $b/2 > z > -b/2$ can be represented as

$$A\xi(z) + C\zeta(z) = A\sin Kz + C\cos Kz, \quad (18)$$

where the wave vectors $K$ and $k_z$ are connected with the relation $K^2/2 = k_z^2/2 + W$. The amplitude of electron wave reflection within this model is defined by the following formula

$$|B|^2 = \frac{(k_z^2 - K^2)^2}{(k_z^2 + K^2)^2 + 4k_z^2 K^2 \cot^2 Kb}. \quad (19)$$

The amplitude of the electron wave passed over the potential well $|1+D|^2$ is connected with (19) as follows

$$1 - |B|^2 = |1+D|^2. \quad (20)$$

As before, the formulas (14) describe the shape of the photoelectron angular distribution and they will be used for numerical calculations.

NUMERICAL CALCULATIONS

*1) S-atomic state*

For the strength of the delta-potential $A$ defining the scattering amplitude (11) to be evaluated, we will assume that the potential of the graphene sheet $U(z)$ is similar to a plane-expanded potential of the fullerene $C_{60}$ shell. Within this assumption the parameter $A$ can be found from the experimental data on the fullerene radius $R$ and the electron affinity energy $I$. According to (Baltenkov, A. S. 1999),

$$A = \beta(1+\coth\beta R)/2, \quad (21)$$

where the parameter $\beta$ is $\beta = (2I)^{1/2}$. For $I = 2.65$ eV and $R = 6.639$ (Tossati, E. et al 1994) the value of the potential strength $A$ is $A = 0.441$.

This value corresponds to spherical potential well of the squared profile with the depth $W$ and the thickness $b$ that are connected with each other so that a bound state with energy $I$ exists in this potential well. In the case of $C_{60}$ it is generally accepted that the fullerene electronic hull has the thickness $b=1.89$ and the depth $W =8.24$ eV (Baltenkov, A. S., et al 2010). It is evident that for the graphene sheet these values of the parameters $A$, $W$ and $b$ should be considered as tentative. At the end of this section we will analyze how the shape of photoelectron angular distribution are modified with variation of these parameters.

Let us consider the case of the *s*-states of the ionized atom. For them the dipole asymmetry parameter is $\beta = 2$ and the function $F(\vartheta,\varphi) = \cos^2 \vartheta_{ek}$. The coordinate system is chosen so that the X-axis is in the same plane as the vectors **e** and Z where the intensity of photoelectron emission is maximal. In this coordinate system the azimuth angle $\varphi$ in the formulas (14) is equal to zero and the photoelectron spectrum shape is defined by the *x*- and *z*-components of the function *S*. The calculation results for the function *S* defining the photoelectron spectrum shape for two polar angles of the **e** vector $\vartheta_e = \pi/2$ and 0 are presented in Figs. 2 and 3. The angular distributions for free atom photoionization are presented in these figures and in the other ones for comparison.

In Fig. 2 the angle $\vartheta_e = \pi/2$ and the function $F(\vartheta,\varphi) = \sin^2 \vartheta$. This function is not equal to zero at the angles $\vartheta = \pi/2$ and $3\pi/2$ and at these points the jumps of the function $S_z$ manifest themselves in the explicit form. The upper spectrum part corresponding to the angular distribution of photoelectrons passing through the potential plane is evidently less than the lower one. With the rise in photoelectron kinetic energy $\varepsilon$ the reflected wave amplitude (15) decreases and the spectrum shape for adatoms is transformed into that corresponding to the free atom.

The curves for the angle $\vartheta_e = 0$ for which the function $F(\vartheta,\varphi)$ is $F(\vartheta,\varphi) = \cos^2 \vartheta$ are presented in Fig. 3. For the angles $\vartheta = \pi/2$ and $3\pi/2$ the function $F(\vartheta,\varphi)$ vanishes and so the jumps in $S_z$ are not explicit. The upper spectrum part corresponding to the angular distribution of photoelectrons passing through the potential plane, as in Fig. 2, is vividly less than the lower one. For **e** || *Z* the spectrum shapes are independent of the azimuth angle $\varphi$ and the 3D-pictures of the photoelectron angular distribution are the figures formed by rotating the curves around the Z-axis. For all the polar angles of the vector **e** the picture of spectrum transformation due to the potential $U(z)$ is qualitatively the same: electron wave reflection leads to enhancing the flux to the lower semi-space and weakening to the upper one. With the rise in photoelectron kinetic energy $\varepsilon$ the adatom spectra are transformed into those corresponding to the free atom.

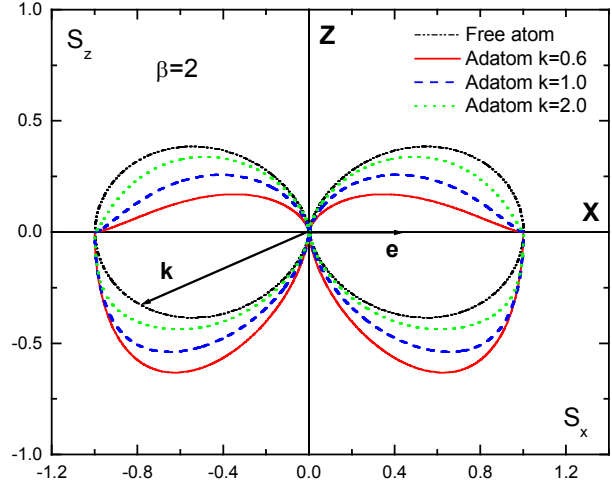

FIG. 2. THE ANGULAR DISTRIBUTIONS OF PHOTOELECTRONS KNOCKED OUT OF THE *S*-ATOMIC SUBSHELLS FOR DIFFERENT ELECTRON MOMENTUMS *K*. THE POLARIZATION VECTOR IS **e** || X. DASH-DOT-DOT BLACK LINE – FREE ATOM; SOLID RED LINE – ATOM + GRAPHENE SHEET FOR *K*=0.6; DASHED BLUE LINE – ADATOM FOR *K*=1.0; DOT GREEN LINE – ADATOM FOR *K*=2.0.

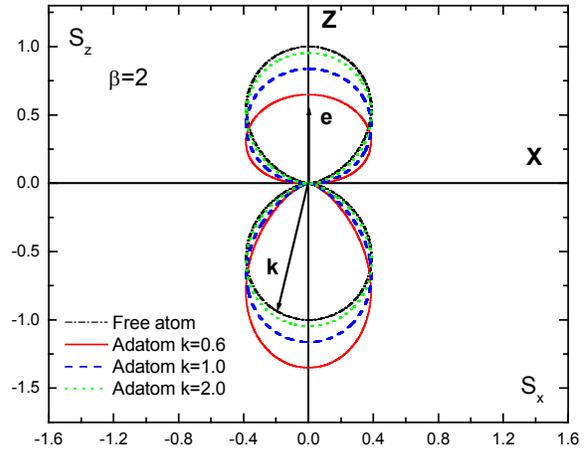

FIG. 3. THE ANGULAR DISTRIBUTIONS OF PHOTOELECTRONS KNOCKED OUT OF THE *S*-ATOMIC SUBSHELLS FOR DIFFERENT ELECTRON MOMENTUMS *K*. THE POLARIZATION VECTOR IS **e** || Z. DASH-DOT-DOT BLACK LINE – FREE ATOM; SOLID RED LINE – ATOM + GRAPHENE SHEET FOR *K*=0.6; DASHED BLUE LINE – ADATOM FOR *K*=1.0; DOT GREEN LINE – ADATOM FOR *K*=2.0.

Let us compare the amplitudes of electron wave reflection in the models of zero and finite thicknesses of the potential well $U(z)$. In the $\delta$-potential model the reflection coefficient is defined by the formula (15). The value of the potential strength $A$ in (15) corresponds to potential well with the above defined parameters $W$ and $b$. Let us consider another several pairs of the parameters $W$ and $b$ meeting the condition of bound state existence in the well (Baltenkov, A. S., et al 2010). The comparison results of the reflection amplitudes (15) and (19) with these parameters are presented in Fig. 4. It is seen that the narrower and deeper the potential well, the wider the electron energy range where the behavior of the reflection amplitudes defining the function S in the both models becomes similar.

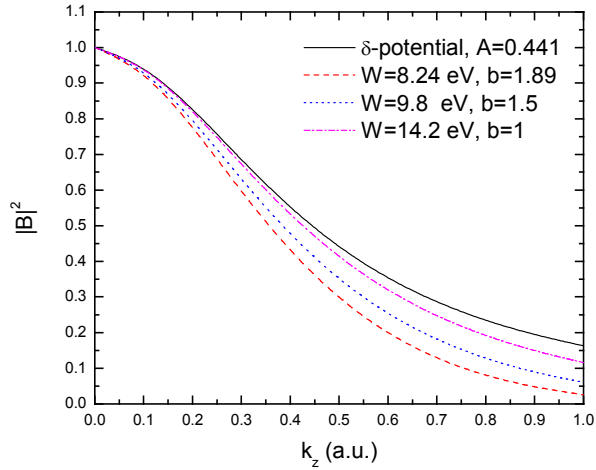

FIG. 4. THE AMPLITUDE OF ELECTRON WAVE REFLECTION IN THE $\delta$-POTENTIAL MODEL AND FOR SOME SQUARED POTENTIALS.

Fig. 5 makes it possible to see the spectrum transformation with changes in the strength of the delta-potential $A$ for fixed electron energy ($k$=0.6). With the rise in the parameter $A$ the reflected wave amplitude (15) goes to unity and the photoelectron flux is almost totally reflected by the potential plane to the lower semi-space.

*2) Non-zero orbital moments*

The photoelectron angular distribution for the atomic shells with nonzero orbital moments is characterized by the dipole asymmetry parameter $\beta$ changeable within the range $-1 \leq \beta \leq 2$. It is possible to see the evolution of the photoelectron spectrum shape for the parameter $\beta$ in the interval $-0.5 \leq \beta \leq 1.0$ in Fig. 6.

The polarization vector **e** here is parallel to the Z-axis. The calculations were performed for two values of electron momentum $k$=0.6 and $k$=1.0. As before, with the rise in electron energy $\varepsilon$ the adatom spectra are transformed into those corresponding to the free atom. The electron angular distribution for parameter $\beta = 0$ corresponds, *per se*, to the picture of the light intensity of the isotropic source (candle) locating near semitransparent mirror when an observer is far from it at the distance significantly exceeding $d$ and so the light source and its reflection in mirror are perceived by him as a point source of light.

As it follows from the figures the existence of the electron waves reflected from the graphene sheet potential $U(z)$ leads to quite observable asymmetry in the angular distribution relative to the plane $z=0$; the less the photoelectron kinetic energy $\varepsilon$ the more pronounced this effect.

For $k$=0.6 the electron wave length is $\lambda = 2\pi/k \approx 10$ while the distance between the carbon atoms in graphene is $a$=1.42Å=2.68 (Abergel, D. S. L., et al 2009). It is this range of electron wave length $\lambda \gg a$ where the developed model is parametrically justified. The experimental observation of this new physical effect is of great interest for photoelectron spectroscopy of atoms localized on graphene structures.

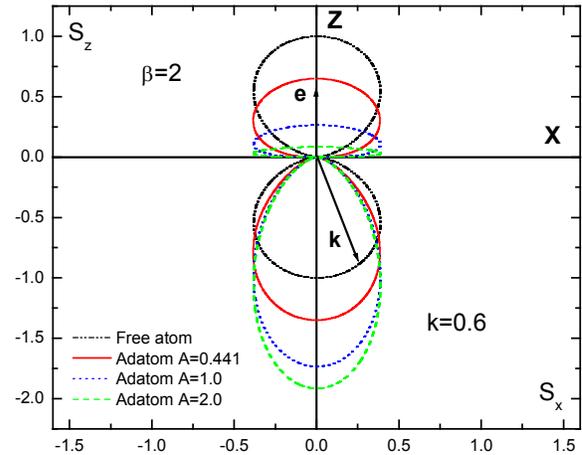

FIG. 5. THE ANGULAR DISTRIBUTIONS OF PHOTOELECTRONS KNOCKED OUT OF THE S-ATOMIC SUBSHELLS FOR DIFFERENT VALUES OF THE PARAMETER $A$ AND FOR FIXED ELECTRON MOMENTUMS $K$=0.6. THE POLARIZATION VECTOR IS $\mathbf{e} \parallel Z$. DASH-DOT-DOT BLACK LINE – FREE ATOM; SOLID RED LINE – ATOM + GRAPHENE SHEET FOR A=0.441; DASHED BLUE LINE – ADATOM FOR $A$=1.0; DOT GREEN LINE – ADATOM FOR $A$=2.0.

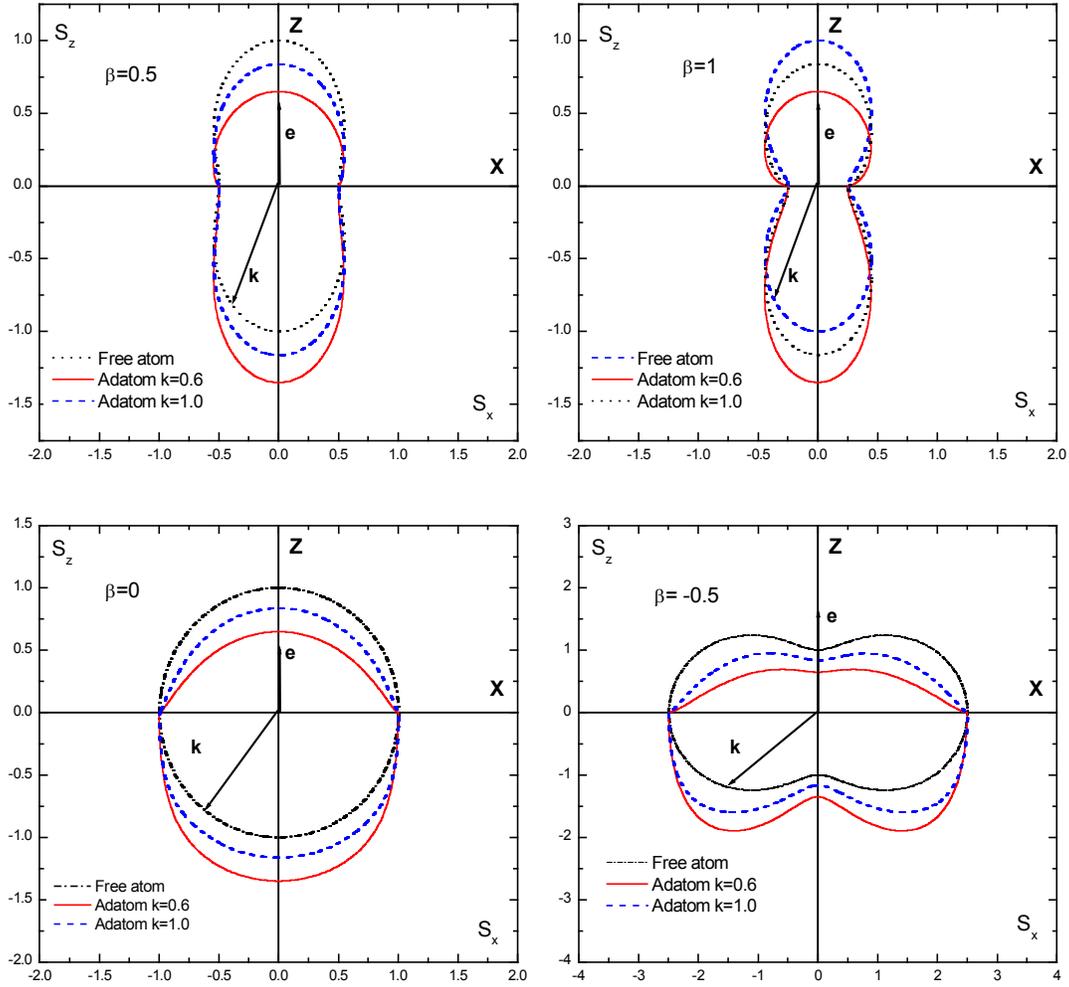

FIG. 6. THE ANGULAR DISTRIBUTIONS OF PHOTOELECTRONS KNOCKED OUT OF THE ATOMIC SUBSHELLS WITH NONZERO ORBITAL MOMENTS FOR DIFFERENT DIPOLE ASYMMETRY PARAMETERS $\beta$ AND DIFFERENT ELECTRON KINETIC ENERGY $\varepsilon = k^2/2$. THE POLARIZATION VECTOR IS $\mathbf{e} \parallel Z$. DASH-DOT BLACK LINES – FREE ATOM; SOLID RED LINES – ATOM + GRAPHENE SHEET FOR $K$=0.6; DASH BLUE LINES – ATOM + GRAPHENE SHEET FOR $K$=1.0. THE 3D-PICTURES OF THE PHOTOELECTRON ANGULAR DISTRIBUTION ARE THE FIGURES FORMED BY THE CURVES ROTATION AROUND THE Z-AXIS.

## ACKNOWLEDGMENT

This work was supported by the Uzbek Foundation Award Ф2-ФА-Ф164.